\documentstyle[12pt,epsfig]{article}
\def\be{\begin{equation}}
\def\ee{\end{equation}}
\def\bc{\begin{center}}
\def\ec{\end{center}}
\def\bea{\begin{eqnarray}}
\def\eea{\end{eqnarray}}

\def\nn{\nonumber}
\def\lp{\lambda^{\prime\prime}}
\def\rp{$R_p$} 
\def\st1{$\tilde{t}_1$}
\def\kkb{$K^0$--$\bar{K}^0$}
\begin{document}
%
%
\begin{titlepage}
\vspace*{-1cm}
\hfill{DFPD-00/TH/40}
\vskip 3.0cm
\begin{center}
{\Large\bf Constraints on R-parity violating \\
stop couplings from flavor physics}
\end{center}
\vskip 1.5  cm
\begin{center}
{\large Pietro 
Slavich}\footnote{e-mail address: slavich@pd.infn.it}
\\
\vskip .1cm
{\em Dipartimento di Fisica, Universit\`a di Padova and \\
INFN, Sezione di Padova, I-35131 Padova, Italy}
\\
\end{center}
\vskip 1.0cm
\begin{abstract}
\noindent
We perform a critical reassessment of the constraints on the R-parity
and baryon number violating (s)top couplings coming from flavor physics.
In particular, we study \kkb\ mixing, including QCD corrections and a class 
of diagrams that were neglected in previous analyses: both effects give
sizeable contributions. We explain how the resulting bounds can be
translated into constraints on individual couplings in the simplest class of
flavor models, based on U(1) horizontal symmetries. We finally discuss the 
implications for the possibility of observing single superparticle production 
at the Tevatron. 

\end{abstract}
\end{titlepage}
\setcounter{footnote}{0}
\newpage


\section{Introduction}
\label{sec1}

In the Minimal Supersymmetric Standard Model (MSSM) it is assumed that
R-parity (\rp) is conserved \cite{fayet}. While the particles of the 
Standard Model are 
even under \rp, their supersymmetric partners are odd, thus the latter can 
only be produced in pairs and they always decay into final states involving 
an odd number of supersymmetric particles. 
Although it considerably simplifies the structure of the MSSM, the 
conservation of R-parity has no firm theoretical justification (for 
a review and references on versions of the MSSM with broken R-parity,
see e.g. \cite{rprev}). 
The most general \rp-violating superpotential that can be written with 
the MSSM superfields contains both lepton number and baryon number violating 
terms. Their simultaneous presence is in fact strongly limited, since 
it would induce fast proton decay. However, it is  possible to allow for
the presence of baryon number violating couplings only \cite{zwirner}, 
coming from the superpotential
\be
\label{superp}
{\cal W}_{\not{B}} = \lp_{ijk}\, \epsilon^{\alpha\beta\gamma} \,\,
U^c_{i\alpha} \,D^c_{j\beta} \,D^c_{k\gamma}\; ,
\ee
where $U^c$ and $D^c$ ($c$ denotes charge conjugation) are the chiral 
superfields associated with SU(2)-singlet antiquarks, 
in a basis where the quark masses are diagonal,
$i,j,k = 1,2,3$ are flavor indices and $\alpha,\beta,\gamma = 1,2,3$ 
are color indices. The color structure of ${\cal W}_{\not{B}}$
implies that the $\lp_{ijk}$ are antisymmetric in the last two indices, 
limiting the number of independent couplings to 9.
 
A number of recent works \cite{berger}-\cite{abraham} studied the 
production of {\em single} supersymmetric
particles at the Fermilab Tevatron through the R-parity and baryon number 
violating couplings of eq. (\ref{superp}) that involve a top (s)quark.
These works were motivated by the consideration that, while the 
\rp\ and baryon number violating couplings $\lp_{1jk}$ which involve an 
up (s)quark are severely constrained by the absence of neutron-antineutron 
oscillations \cite{zwirner,goity} and double nucleon decay into kaons 
\cite{masiero}, the limits on the couplings that involve a top (s)quark 
are believed to be much weaker, of order $\lp_{3jk}<1$ for sfermion masses 
heavier than 100 GeV (see e.g. \cite{fermilab} and references therein). 
Moreover, in most scenarios of supersymmetry breaking it is reasonable to 
expect that the third-generation squarks are lighter than the other 
sfermions, due to
the effect of their large Yukawa couplings on the renormalization group 
evolution of the masses between the ultraviolet cutoff scale 
of the MSSM and the weak scale. 
Thus, if the couplings $\lp_{3jk}$ are not much smaller than 1, 
and the third-generation squarks are not too heavy, it should be possible 
to observe single stop production in the interaction of two down-type 
antiquarks, or single sbottom production in the decay of a top quark.

In view of this situation, it is extremely interesting to understand 
how the constraints on the \rp-violating (s)top couplings that can be 
derived from low-energy experimental data compare with the bound 
$\lp_{3jk} < 1$ considered in Refs. \cite{berger}-\cite{abraham}. 
In this note we will study the constraints that can be derived from 
flavor physics.
In particular, in section \ref{sec2} we will perform a careful study of 
the limits coming from \kkb\ mixing, including QCD corrections and a class 
of diagrams that were neglected in previous analyses 
\cite{masiero}-\cite{white}: both effects give sizeable contributions. 
We will show that \kkb\ mixing places an upper bound of order (few)$\times
10^{-2}$ on the combination 
$|\lp_{313} \lambda^{\prime\prime\,*}_{323}|^{1/2}$ 
of \rp-violating (s)top couplings. For degenerate couplings, 
$\lp_{312} = \lp_{313} = \lp_{323} \equiv \lp$, 
this is significantly stronger than the bounds usually considered in the 
literature. 
In section \ref{sec3} we will consider a more refined way of translating 
the bounds coming from \kkb\ mixing into bounds on the individual
couplings $\lp_{3jk}$: it makes use of the simplest models of flavor,
those based on U(1) horizontal symmetries, to correlate the size of the 
different couplings.
Finally, in section \ref{sec4} we will apply our analysis to the specific
scenarios considered in Refs. \cite{berger}-\cite{abraham}, and we will 
show that \kkb\ mixing strongly constrains the possibility of observing 
single superparticle production at the Tevatron, not only under the rough 
assumption of degenerate couplings, but also assuming the most general 
hierarchies allowed by realistic U(1) flavor symmetries.

Before addressing our main points, we comment on some bounds that were 
previously discussed in the literature. The R-parity and baryon number 
violating interactions affect, at the one-loop level, the coupling of 
the $Z$ boson to the right-handed down-type quarks. The upper bounds quoted 
in Refs. \cite{berger}-\cite{abraham} come from the measurement of 
$R_l \equiv \Gamma(Z \rightarrow$ hadrons)/$\Gamma(Z \rightarrow l^+ l^-)$
\cite{bhatta}.  
However, in order to derive a reliable bound on the couplings $\lp_{3jk}$,
the contributions of both \rp-violating and \rp-conserving MSSM interactions 
to the $Z$-peak observables must be taken into account \cite{yang}. 
A recent analysis of the whole set of hadronic $Z$-peak observables 
\cite{takeuchi} suggested that the couplings $\lp_{3j3}$ are ruled out 
at 95\% confidence level. The reason is that, while the data ask for a 
positive shift in the coupling of the right-handed $b$ quarks to the $Z$ 
boson with respect to its Standard Model value, the $\lp$ corrections 
shift this coupling in the negative direction. However, the interpretation
of these results appears to be controversial: it was shown in \cite{takeuchi} 
that the bounds on $\lp_{3j3}$ coming from the $Z$-peak data are much weaker 
if one performs a Bayesian statistical analysis instead of the classical one.

The experimental bound on neutron-antineutron oscillations was used
\cite{goity} to set limits on the \rp-violating couplings $\lp_{1jk}$ which
involve an up (s)quark. In a subsequent paper \cite{keung}, a similar analysis
was performed on a class of diagrams that also contribute to 
neutron-antineutron oscillations, but allow to set limits on the couplings  
$\lp_{3jk}$. Contrary to what claimed by the authors of \cite{keung}, 
we have found that such limits are not competitive with the ones that can 
be derived from flavor physics.

The authors of \cite{carlson} studied the contribution of the \rp-violating 
terms to some rare decays of the $B^+$ meson, namely 
$B^+ \rightarrow \bar{K}^0 K^+$ and $B^+ \rightarrow \bar{K}^0 \pi^+$. 
In order to reduce the theoretical 
uncertainties, they considered the ratio of the partial width of the 
rare decay to the partial width of the decay $B^+ \rightarrow K^+ \, J/\psi$, 
which proceeds unsuppressed in the Standard Model. 
The limits on the \rp-violating couplings that can be derived in this
way are less stringent than those coming from \kkb\ mixing, but still one
order of magnitude below those assumed in \cite{berger}-\cite{abraham}. 
Using the formulae of Ref. \cite{carlson}, with updated values for the
experimental upper bounds on the branching ratios and for the CKM matrix
elements \cite{pdg}, and assuming that the lightest stop mass eigenstate is
approximated by $\tilde{t}_R$, we obtained:
\begin{equation}
\label{roy}
|\lp_{312} \lambda^{\prime\prime\,*}_{323}|
< 3.0 \times 10^{-3} \; \frac{m_{\tilde{t}_R}^2}{m_W^2}\; ,  \hspace{1.5cm}
|\lp_{312} \lambda^{\prime\prime\,*}_{313}|
< 3.9 \times 10^{-3} \; \frac{m_{\tilde{t}_R}^2}{m_W^2}\; .
\end{equation} 
As an example we can consider the representative parameter choice 
recently used for the discussion of single stop production at the
Tevatron (see section \ref{sec4} for the details concerning this 
scenario). In the case of degenerate couplings 
$\lp_{312} = \lp_{313} = \lp_{323} \equiv \lp$, and taking into account the 
mixing in the stop sector, we obtained the upper limit $\lp < 0.15-0.24$ 
for a light stop mass  ranging between 180 and 325 GeV. 

\section{Improved constraints from \kkb\ mixing}
\label{sec2}

Bounds on the \rp-violating couplings can be derived from \kkb\ mixing 
\cite{masiero}-\cite{white}. Flavor-changing neutral currents 
in SUSY models can arise in a ``direct'' way, when the 
flavor violation occurs through flavor violating vertices in the diagrams, 
or in an ``indirect'' way, due to the existence of non diagonal sfermion 
masses in the basis where the fermion masses are diagonal. In minimal 
supergravity scenarios, where the soft mass matrices at the GUT scale are 
flavor diagonal, non diagonal squark masses are generated by flavor violating 
couplings through the renormalization group equations. However, as shown 
in \cite{white}, their contribution to \kkb\ mixing can be neglected.
Thus, in the following we will assume that the quark and squark mass 
matrices are diagonalized by the same rotations.

The diagrams that give the dominant contributions to \kkb\ mixing 
in minimal supergravity scenarios with R-parity violation are shown in fig. 
\ref{fig-box}.
The most general $\Delta S = 2$ effective Lagrangian can be written as:  
\be
\label{lagr-KK}
{\cal L}_{\rm eff}^{\Delta S = 2} =  \sum_{i=1}^5 C_i Q_i + \sum_{i=1}^3 
\tilde{C}_i\tilde{Q}_i  \; ,
\ee
where the four-fermion operators $Q_i$ and $\tilde{Q}_i$ are defined 
as in \cite{ciuchini}. The operators relevant to this analysis are 
$Q_1 = \bar{d}_L^{\alpha} \gamma_{\mu}s_L^{\alpha}\;
\bar{d}_L^{\beta} \gamma^{\mu}s_L^{\beta}$ ($\alpha,\beta$ are color indices), 
coming from the Standard Model diagram (fig. \ref{fig-box}a),
$\tilde{Q}_1 = \bar{d}_R^{\alpha} \gamma_{\mu}s_R^{\alpha}\;
\bar{d}_R^{\beta} \gamma^{\mu}s_R^{\beta}$, 
coming from the diagrams with four $\lp$ couplings (fig. \ref{fig-box}b-c), 
and $Q_4  =  \bar{d}_R^{\alpha} s_L^{\alpha}\;
\bar{d}_L^{\beta} s_R^{\beta},\; Q_5  =  \bar{d}_R^{\alpha}  s_L^{\beta}\;
\bar{d}_L^{\beta}  s_R^{\alpha}$, coming from the diagrams with two CKM and 
two $\lp$ couplings (fig. \ref{fig-box}d-e).
The corresponding coefficients $C_i$ are evaluated at a common scale $M_S$, 
where the supersymmetric particles are integrated out. We computed all
the coefficients that are relevant to the case under consideration,
including the contributions of charginos and charged Higgs that were 
neglected in earlier analyses \cite{masiero}-\cite{white}: 
their explicit expressions are given in the Appendix.

The contribution of the effective Lagrangian ${\cal L}^{\Delta S = 2}$ 
to the $K_S$--$K_L$ mass difference $\Delta m_K$ is related to the matrix 
element $\langle K^0 | {\cal L}_{\rm eff}^{\Delta S = 2} | \bar{K}^0 \rangle$. 
The coefficients $C_i$ must be evolved from the scale $M_S$, which 
is of order of the masses of the supersymmetric particles,  down to
some hadronic scale $\mu_h$ (e.g. 2 GeV) at which the matrix 
element can be evaluated.
Moreover, the long-distance hadronic processes give  contributions to
the matrix elements  $\langle K^0 | Q_i | \bar{K}^0 \rangle$ that cannot
be evaluated perturbatively, and are parametrized by ``bag factors'' $B_i$
(for the explicit definitions see \cite{ciuchini}). We have calculated the
contribution of the \rp-violating couplings to $\Delta m_K$, using the 
NLO QCD evolution of the coefficients $C_i$ and the lattice calculations
for the $B_i$ presented in \cite{ciuchini}.
It is interesting to note that the main effect of the QCD corrections, 
which were neglected in earlier analyses, is a sizeable enhancement 
of the coefficient $C_4$.
 
Due to the large uncertainties that affect the theoretical evaluation 
of $\Delta m_K$ in the Standard Model 
(see e.g. \cite{nierste} and references therein), a 
conservative limit on the \rp-violating couplings can be 
derived by requiring that the contribution to $\Delta m_K$ of the diagrams 
shown in fig. \ref{fig-box}b-e is not larger than the experimental value 
$\Delta m_K^{\rm exp} = (3.489 \pm 0.008) \times 10^{-15}$ GeV \cite{pdg}. 
When all the contributions of the diagrams shown in fig. \ref{fig-box}b-e
are taken into account, the resulting limit on the product 
$\lp_{313}\lambda^{\prime\prime\,*}_{323}$ depends in a nontrivial way on 
the spectrum of sparticle masses and mixing angles. As a first step, we 
derived an upper bound on the \rp-violating couplings by performing a general 
scan over the parameter space of the MSSM at the weak scale. 
According to the standard lore on the hierarchy problem, we varied the soft 
squark masses $m_{Q_3}$, $m_{U_3}$ and $m_{D_3}$, 
the gaugino mass $M_2$ and the charged Higgs mass $m_{H^+}$ between 100 
GeV and 1 TeV, the squark trilinear couplings $A_t,A_b$ and the $\mu$ 
parameter between $-1$ TeV and 1 TeV, and $\tan\beta$ between 1 and 50, 
with the constraint that the mass spectrum of superpartners is not already 
excluded by direct searches. The result is that, for any choice of the 
MSSM parameters, the upper limit on the \rp-violating couplings is
of order $|\lp_{313}\lambda^{\prime\prime\,*}_{323}|^{1/2} <0.033$ or lower
(the weakest bound is obtained when $m_{U_3}\,, m_{D_3}$ and $|\mu|$ get their
maximal value). This is significantly stronger than the upper limit 
$|\lp_{313}\lambda^{\prime\prime\,*}_{323}|^{1/2} < 0.12$ 
quoted in \cite{carlson} for the case $m_{\tilde{b}_R} = 1 $ TeV, 
$m_{\tilde{t}_R} = 900$ GeV. However, the authors of \cite{carlson} neglected 
in their analysis the contribution of the diagram with chargino exchange
shown in fig. \ref{fig-box}e, as well as the effect of squark mixing and 
QCD corrections. 

\section{Constraints from flavor symmetries}
\label{sec3}

As noticed above, flavor changing processes  allow to set bounds only on 
products of two different $\lp_{3jk}$ couplings. 
To discuss the implications for 
single superparticle production at colliders, however, it would be 
interesting to translate these bounds into bounds on the individual 
couplings. Two extreme but simple-minded choices that are often made
in the literature consist in assuming either degenerate couplings
or only one non-vanishing coupling: in the first case,
$\lp_{312} = \lp_{313} = \lp_{323} \equiv \lp$,
the limit from \kkb\ mixing is simply $\lp < 0.033$ or
lower, whereas in the second case the only non-vanishing coupling 
is unconstrained. However, none of these two choices is natural,
as can be easily realized considering flavor models based on
horizontal symmetries. As a representative case we shall concentrate
here on the class of models based on a U(1) flavor symmetry, but similar
considerations could be made for all other realistic flavor models. In 
abelian flavor models, the Yukawa couplings for up-type and down-type 
quarks are of order 
$Y^U_{ij} \sim \epsilon^{h_2 +q_i+u_j}$ and 
$Y^D_{ij} \sim \epsilon^{h_1 +q_i+d_j}$ respectively, while the R-parity 
and baryon number violating couplings are of order 
$\lp_{ijk} \sim \epsilon^{u_i+d_j+d_k}$, 
where  $\epsilon \sim 0.2$ is a parameter of the order of the Cabibbo angle
and $h_i,\,q_i,\,u_i$ and $d_i$ denote the charges of the corresponding
MSSM superfields (in the basis of interaction eigenstates) under the 
additional U(1) symmetry. It has been 
shown \cite{dudas}-\cite{choi} that the correct pattern of quark masses and 
mixing angles can be reproduced only with the following two sets of charge 
assignments:
\be
{\rm (I)} \hspace*{1.1cm}
(q_{13},q_{23}) = (3,2),\;\;\;\;\; 
(u_{13},u_{23}) = (5,2),\;\;\;\;\; 
(d_{13},d_{23}) = (1,0),
\ee
\be
{\rm (II)} \hspace*{1cm}
(q_{13},q_{23}) = (-3,2),\;\;\; 
(u_{13},u_{23}) = (11,2),\;\;\; 
(d_{13},d_{23}) = (7,0)
\ee
where $q_{ij} \equiv q_i - q_j$ etc. In the basis of interaction eigenstates,
the hierarchy among the $\lp_{3jk}$ couplings follows the pattern: 
\be
\label{pattern}
\lp_{312} \approx \lp_{313} \approx \kappa\,\lp_{323},
\ee 
where $\kappa = \epsilon$ in case I (i.e. a mild hierarchy) and 
$\kappa = \epsilon^7$ in case II (a strong hierarchy). 
However, the hierarchy is attenuated in the basis of eq. (\ref{superp}), 
where the quarks are mass eigenstates. 
Following Refs. \cite{hall,dudas,ramond} the quark mixing matrices can be 
evaluated at leading order in $\epsilon\,$ upon diagonalization of the 
Yukawa matrices.
We found that in the basis of quark mass eigenstates the hierarchy 
among the $\lp_{3jk}$ couplings follows the pattern of eq. (\ref{pattern}), 
with $\kappa = \epsilon\,$ in case I and $\kappa = \epsilon^3$ in case II.
In the presence of such hierarchy, the upper limit on the unsuppressed
coupling $\lp_{323}$ coming from \kkb\ mixing is rescaled by a factor
of $\kappa^{-1/2}$ with respect to the case of degenerate couplings, while
the upper limit on the suppressed couplings $\lp_{312}$ and $\lp_{313}$ 
is rescaled by a factor of $\kappa^{1/2}$: the resulting bounds are 
$\lp_{323} < 0.07 \,,\; \lp_{312} \approx \lp_{313} < 0.015$ in case I and
$\lp_{323} < 0.37 \,,\; \lp_{312} \approx \lp_{313} < 0.003$ in case II.
Thus, even in the less stringent situation, the upper bound on the 
largest R-parity violating (s)top coupling is at most of order 
(few)$\times 10^{-1}$, i.e. one order of magnitude below the bounds assumed 
in Refs. \cite{berger}-\cite{abraham}. 

\section{Implications for single superparticle production}
\label{sec4}
\begin{figure}
\begin{center}
\epsfig{figure=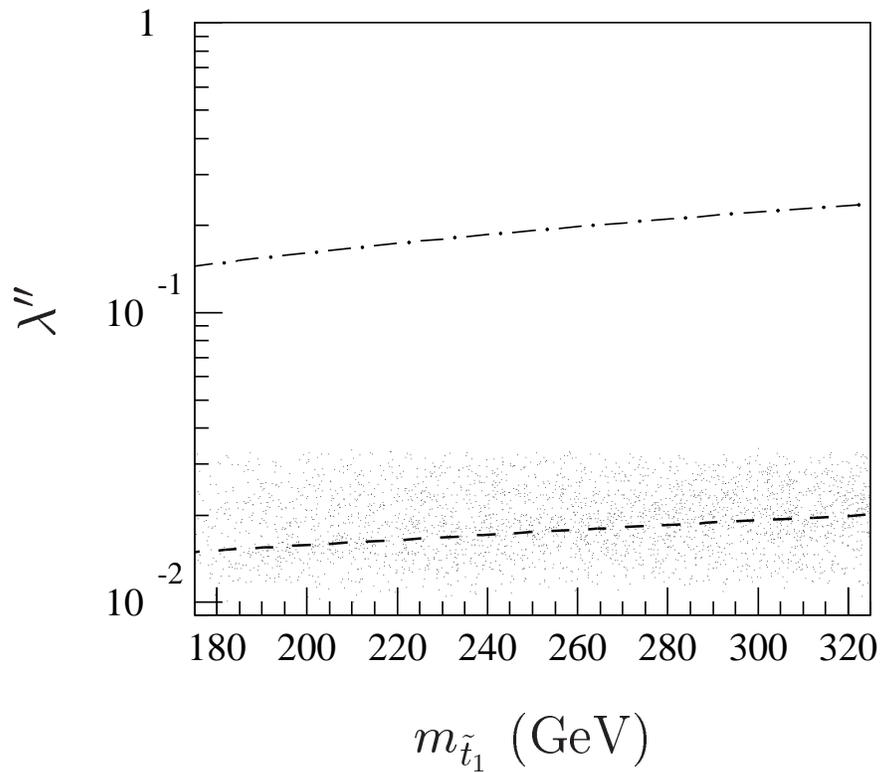,width=14cm}
\end{center}
\caption{Upper limits on $\lp$ coming from \kkb\ mixing (dashed line) 
and $B$ rare decays (dot-dashed line) in the scenario relevant to
single stop production, with the choice of parameters considered in 
\cite{berger}. The scattered points are the limits from \kkb\ mixing that 
result from a wide scan over the MSSM parameter space.}
\label{fig-limits}
\end{figure}

We are now ready to discuss the implications of our results for single 
superparticle production at colliders. 
In \cite{berger} the production of single 
top squarks via \rp\ violation in $p \bar{p}$ collisions at the 
Tevatron was studied. If some of the couplings $\lp_{3jk}$ are greater than
$10^{-2}$ or so, the rate for the production of a single light stop \st1\ in 
$p \bar{p}$ collisions at $\sqrt{S} = 2$ TeV may exceed the 
rate for stop-antistop pair production, due to the greater phase space 
available. Thus, \rp\ violation could be the favorite scenario for the 
observation of supersymmetric particles at the Tevatron.
The authors of \cite{berger} considered a supergravity-inspired scenario,
where at the Grand Unified Theory (GUT) scale the common gaugino mass is 
$m_{1/2} = 150$ GeV, the scalar trilinear coupling is $A_0 = -300$ GeV 
and the common scalar mass $m_0$ is varied in a range between 
50 and 500 GeV. The ratio of the Higgs vacuum expectation 
values is chosen to be $\tan\beta = 4$ and the Higgs mass parameter $\mu$, 
whose absolute value is fixed by electroweak symmetry breaking, is chosen
to be positive. The three \rp-violating (s)top couplings  were taken to be 
degenerate, $\lp_{312} = \lp_{313} = \lp_{323} \equiv \lp$.
The signal coming from single stop production in $p\bar{p}$ collisions,
followed by the \rp-conserving decay \st1$\rightarrow b + 
\tilde{\chi}^+_1$, with $\tilde{\chi}^+_1 \rightarrow l + \nu + 
\tilde{\chi}^0_1$, was considered in \cite{berger} together with the Standard 
Model background. The conclusion was that,
for $ 180 < m_{\tilde{t}_1} < 325$ GeV and $\lp > 0.02-0.06$,
it should be possible to discover the top squark at run II of the Tevatron, 
otherwise the limit on the \rp-violating couplings could be lowered to 
$\lp < 0.01-0.03$ at 95\% confidence level. Moreover, existing 
data from run I should allow for a reduction of the limit to $\lp < 0.03-0.2$
for $ 180 < m_{\tilde{t}_1} < 280$ GeV.
We computed the upper bound on $\lp$ coming from \kkb\ mixing
in the minimal supergravity scenario 
considered by the authors of \cite{berger}, for the same choice of 
parameters.  The resulting limits are of order $\lp < 0.015-0.020$ for a 
light stop mass ranging between 180 and 325 GeV, as shown in fig. 
\ref{fig-limits}. In the same figure, the limits resulting from the
general scan over the MSSM parameters described above are shown, and 
it can be seen that they never exceed $\lp < 0.033$. The (weaker) limits
coming from rare decays of the $B^+$ boson (see section \ref{sec1}) 
are also shown. In summary, under the assumption of degenerate \rp-violating 
(s)top couplings, the discovery of single stop production via R-parity 
violation at the Tevatron turns out to be unlikely.

In the presence of a hierarchy among the couplings such as that of 
eq. (\ref{pattern}), suggested by U(1) flavor models, the lower limit 
for single stop discovery on the unsuppressed coupling $\lp_{323}$
must be rescaled by a factor 
of $(0.95\, \kappa^2 + 0.05)^{-1/2}$ with respect to the case of degenerate 
couplings, taking into account that the relative contribution of
$\lp_{323}$ to the total cross section $p\bar{p} \rightarrow \tilde{t}_1$ 
is roughly $5\%$. On the other side, as discussed in section \ref{sec3}, the 
upper limit on $\lp_{323}$ coming from \kkb\ mixing is rescaled by a factor 
of $\kappa^{-1/2}$. In case I ($\kappa = \epsilon\,$) the upper limit  from 
\kkb\ mixing becomes $\lp_{323} < 0.034 -0.045$, 
while the lower limit for single stop discovery becomes 
$\lp_{323} > 0.07 - 0.20$: no room for single stop discovery at the 
Tevatron is left. 
In case II ($\kappa = \epsilon^3\,$) the upper limit  from \kkb\ mixing 
becomes $\lp_{323} < 0.17-0.22$, while the lower limit for single stop 
discovery becomes $\lp_{323} >0.09-0.27$. Although single stop discovery 
at the Tevatron cannot be excluded in this case, most of the parameter 
space considered in \cite{berger} is ruled out. 

A very similar discussion can be made concerning the results of \cite{chai},
where the production of single gluinos at hadron colliders,
$p\bar{p}\, (pp) \rightarrow \bar{t} \tilde{g}\,(t \tilde{g})$, was studied.
Such process occurs through the exchange of a virtual squark, produced 
in the R-parity violating interaction between two quarks. The authors of 
\cite{chai} considered a supergravity-inspired scenario where at the 
GUT scale $m_0 = 1$ TeV, $A_0 = -1$ TeV, $\tan\beta = 10$, 
$\mu > 0$ and $m_{1/2}$ is varied between 120 and 400 GeV, 
i.e. a region of the parameter space where the squarks are heavier than
the gluino (with the possible exception of the light stop). The \rp-violating
stop couplings were taken to be degenerate, $\lp_{3jk} \equiv \lp$. 
The conclusion was that, if the gluino is lighter than 400 GeV and 
$\lp$ is of order 1, it will be possible to detect single gluino 
production at the Run II of the Tevatron. At the LHC the process can 
potentially be seen with the gluino lighter than 1 TeV and $\lp$ of order
0.01. We computed the upper bound on $\lp$ coming from \kkb\ mixing,
with the choice of parameters considered in \cite{chai}.
The result is that $\lp < 0.028-0.035$ for a gluino mass ranging between 
350 GeV and 1 TeV. Thus, under the assumption of degenerate $\lp_{3jk}$ 
couplings, the discovery of single gluino production at the Tevatron is 
unlikely, while room is left for discovering the same process at the LHC. 
The same conclusion holds in the presence of a moderate hierarchy among the 
couplings, such as the one suggested by the U(1) flavor models of section
\ref{sec3}.

\begin{figure}
\begin{center}
\epsfig{figure=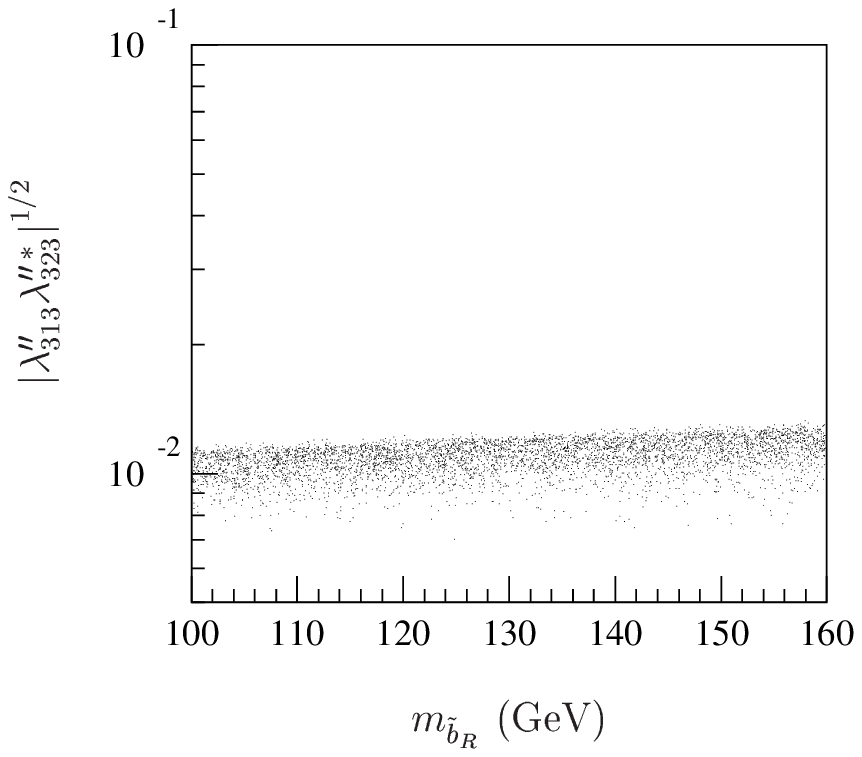,width=14cm}
\end{center}
\caption{Upper limits on
$|\lp_{313}\lambda^{\prime\prime\,*}_{323}|^{1/2}$ 
coming from \kkb\ mixing, in the scenario relevant to
\rp-violating top decays \cite{abraham}. The scattered points are the 
limits that result when $M_2 = 100$ GeV, $\tan\beta = 1$,
$\mu = -200$ GeV and the other MSSM parameters are varied as 
described in the text.}
\label{fig-limits2}
\end{figure}

In \cite{abraham} the decay $t \rightarrow \bar{b}\,\bar{d}\,\tilde{\chi}^0_1$ 
was studied. If the light sbottom has a non-negligible right-handed component,
the top quark may undergo the \rp-violating decay 
$t \rightarrow \bar{\tilde{b}}_1\,\bar{d}$ through the coupling $\lp_{313}$, 
followed by the \rp-conserving decay $\bar{\tilde{b}}_1\rightarrow 
\bar{b}\,\tilde{\chi}^0_1$. The authors of \cite{abraham} considered a scenario
where at the weak scale $\mu = -200$ GeV, $\tan\beta = 1$ and
$M_2 = 100$ GeV. With this choice of the parameters the masses of the 
lightest chargino and neutralino are 
$m_{\tilde{\chi}^+_1} = 120$ GeV and $m_{\tilde{\chi}^0_1} = 55$ GeV,
respectively. The conclusion of \cite{abraham} was that, if the sbottom 
$\tilde{b}_1$ is lighter than 160 GeV and\footnote{Our normalization 
of the couplings $\lp_{ijk}$ (eq. \ref{superp}) differs by a factor of 2 
from the one used in \cite{abraham}.} $\lp_{313} > 0.5$, it will be possible 
to observe the \rp-violating top  decay at the Run II of the Tevatron, 
while the observation 
at the LHC will be possible for $\lp_{313} > 0.1$. The same conclusion 
holds for the decay $t \rightarrow \bar{b}\,\bar{s}\,\tilde{\chi}^0_1$, driven
by the \rp-violating coupling  $\lp_{323}$. We computed the upper limits on 
the combination $|\lp_{313}\lambda^{\prime\prime\,*}_{323}|^{1/2}$, 
using for $M_2,\, \mu$ and $\tan\beta$ the same values as in \cite{abraham}, 
and varying the other MSSM parameters in the way described in section 
\ref{sec2}. The results are shown in fig. \ref{fig-limits2} as a function of 
the right sbottom mass $m_{\tilde{b}_R}$ (with our choice of $\mu$ and 
$\tan\beta$ the mixing in the sbottom sector is negligible). It can be 
noticed that, when the right sbottom is lighter than 160 GeV, the upper 
bounds on the \rp-violating couplings are of order
$|\lp_{313}\lambda^{\prime\prime\,*}_{323}|^{1/2} < 0.013$ 
or lower. Thus we conclude that, if the couplings $\lp_{313}$ and $\lp_{323}$ 
are degenerate (or if there is a hierarchy between the couplings such as 
the one 
suggested by U(1) flavor models), the discovery of \rp-violating top decays 
is unlikely. In any case, we cannot expect to observe both the decays  
$t \rightarrow \bar{b}\,\bar{d}\,\tilde{\chi}^0_1$ and 
$t \rightarrow \bar{b}\,\bar{s}\,\tilde{\chi}^0_1$.

\section{Conclusions}
\label{sec5}

In summary, we have improved the existing bounds on the \rp-violating 
couplings $\lp_{3jk}$, showing that they are typically more stringent than 
those assumed in \cite{berger}-\cite{abraham}. For the bounds coming from 
\kkb\ mixing we have included QCD corrections and a class of diagrams 
that were neglected in earlier analyses \cite{masiero}-\cite{white}: both 
effects give sizeable contributions. We have discussed a way of translating
the resulting bounds into constraints on the individual couplings, making 
use of the simplest class of flavor models based on U(1) horizontal
symmetries. Finally, we have shown that our bounds put severe constraints
on the possibility of observing single superparticle production via R-parity 
violation at the Tevatron.

\vspace*{0.5cm}

\section*{Acknowledgements}
The author thanks Fabio Zwirner for many useful discussions,
and Guido Martinelli for providing the revised version of the 
`magic numbers' of \cite{ciuchini} before publication.

\vspace*{0.5cm}

\section*{Appendix}
\begin{figure}
\begin{center}
\epsfig{figure=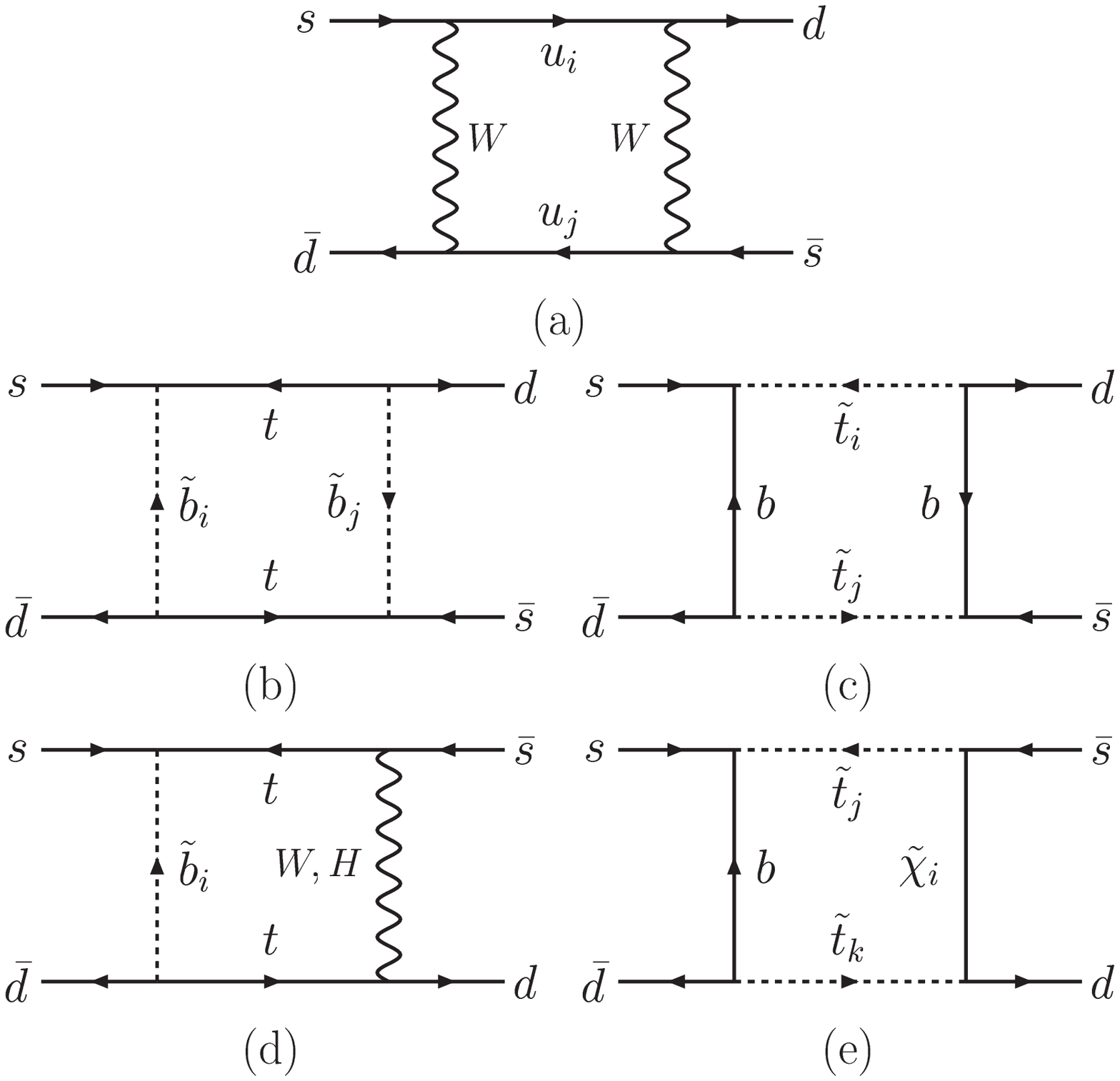,width=14cm}
\end{center}
\caption{Standard Model diagram (a) and diagrams with \rp-violating couplings
(b-e) that give the dominant contributions to \kkb\ mixing. 
The arrows indicate the flow of baryon number.}
\label{fig-box}
\end{figure}

We have calculated the contributions to \kkb\ mixing coming from the 
full set of diagrams with \rp-violating (s)top couplings 
$\lambda^{\prime\prime}_{3jk}$. In particular, the contribution of 
the diagram in fig. \ref{fig-box}e, which was neglected in earlier 
analyses \cite{masiero}-\cite{white}, turns out to be sizeable.
The calculation has been performed in the basis where the quark masses are 
diagonal, and all the flavor changing squark mass insertions have been 
neglected. We have also checked that in the scenarios considered in 
\cite{berger}-\cite{abraham} the contributions coming from the R-parity 
conserving sector of the MSSM (i.e. from diagrams with quarks and 
charged Higgs or squarks and charginos in the internal lines) are 
negligible. The coefficients $C_i$ that appear in eq. (\ref{lagr-KK}) are:
\bea
C_1 & = & \!\sum_{i,j=1}^3 \;\frac{g^4}{128 \pi^2}\,
K^*_{i1}\,K_{i2}\,K^*_{j1}\,K_{j2}\, m^2_{u_i}\,m^2_{u_j}\,
\biggr[ I_0 + 2\, I_2/m_W^2 + I_4/4\,m_W^4 \biggr]
(m^2_{u_i},m^2_{u_j},m_W^2,m_W^2) \nn\\ 
\label{coeff1}
&&\\
\tilde{C}_1 & = & \!\sum_{i,j=1}^2 \;\frac{1}{4 \pi^2}\,
|\lambda^{\prime\prime}_{313}\,\lambda^{\prime\prime\,*}_{323}|^2\,
\biggr[
(O^t_{i2} O^t_{j2})^2 \, 
I_4(m_b^2,m_b^2,m^2_{\tilde{t}_i},m^2_{\tilde{t}_j}) 
+ (O^b_{i2} O^b_{j2})^2 \, 
I_4(m^2_{\tilde{b}_i},m^2_{\tilde{b}_j},m^2_{t},m^2_{t})
\biggr]\nn\\
\label{coeff1t}
&&\\
C_5 & = & \sum_{i=1}^2 \;\;\;\frac{g^2}{4 \pi^2}\,
\lambda^{\prime\prime}_{313}\,\lambda^{\prime\prime\,*}_{323}\,
(O^b_{i2})^2 \, K^*_{31}\, K_{32} \, m^2_t \, 
\biggr[ I_2(m^2_{\tilde{b}_i},m_W^2,m^2_t,m^2_t)\, + \nn\\ 
&& \hspace{2.5cm}  \frac{1}{4\, m_W^2}\, 
I_4(m^2_{\tilde{b}_i},m_W^2,m^2_t,m^2_t)\, +\, 
\frac{1}{4\, m_W^2 \tan^2\beta}\,
I_4(m^2_{\tilde{b}_i},m_{H^{+}}^2,m^2_t,m^2_t)\,  \biggr] \nn\\  
& + & \!\!\! \sum_{i,j,k=1}^2 \; \frac{g^2}{8 \pi^2}\,
\lambda^{\prime\prime}_{313}\,\lambda^{\prime\prime\,*}_{323}\,
O^t_{j2}\,O^t_{k2} \, K^*_{31} \, K_{32} \,
\left[V^*_{i1}\, O^t_{j1} - \frac{m_t}{\sqrt{2}\, m_W \sin\beta} \,
V^*_{i2} \, O^t_{j2} \right] \times \nn\\
\label{coeff5}
&& \hspace{2.5cm}
\left[V_{i1}\, O^t_{k1} - \frac{m_t}{\sqrt{2}\, m_W \sin\beta}\, 
V_{i2}\, O^t_{k2} \right] \; 
I_4(m^2_b,m^2_{\tilde{\chi}^{+}_i},m^2_{\tilde{t}_j},m^2_{\tilde{t}_k})
\eea
and $C_4 = - C_5$. In the above equations, $K_{ij}$ are the CKM 
matrix elements, $O^t_{ij}$ and $O^b_{ij}$ are the left-right mixing 
matrices of the stop and sbottom sectors, and $V_{ij}$ is the mixing matrix 
of positive charginos as defined in \cite{haberkane}. 
The functions $I_n$ result from integration over the Euclidean momentum 
$\bar{k}$ of the four particles circulating in the loop:
\be
\label{def-I}
I_n(m^2_1,m^2_2,m^2_3,m^2_4) = \int_0^{\infty}
\frac{\bar{k}^n \; d\bar{k}^2}
{(\bar{k}^2+m_1^2)(\bar{k}^2+m_2^2)(\bar{k}^2+m_3^2)(\bar{k}^2+m_4^2)}
\ee
In the numerical study of the minimal supergravity scenario, the masses of 
the supersymmetric particles and the mixing angles at the weak scale 
have been calculated with ISAJET \cite{isajet}, and the common scale 
$M_S$ has been chosen as the geometrical mean of squark and chargino masses. 

\vspace*{0.5cm}

\newpage

\newpage

\newpage


\begin{thebibliography}{99}
\bibitem{fayet}
G. R. Farrar and P. Fayet, Phys. Lett. B 76 (1978) 575.
\bibitem{rprev}
H. Dreiner, in {\em Perspectives on Supersymmetry}, ed. by G. L. Kane, 
World Scientific, Singapore (1998), p. 462-479, preprint hep-ph/9707435. 
\bibitem{zwirner}
F. Zwirner, Phys. Lett. B 132 (1983) 103.
\bibitem{berger}
E. L. Berger, B. W. Harris and Z. Sullivan, Phys. Rev. Lett. 83 (1999) 4472.
\bibitem{fermilab}
B. Allanach {\em et al.}, to appear in the {\em Proceedings of the Workshop 
on Physics at Run II - Supersymmetry/Higgs}, Fermilab (1998), 
preprint hep-ph/9906224.
\bibitem{plehn}
T. Plehn, Phys. Lett. B 488 (2000) 359.
\bibitem{chai}
M. Chaichian, K. Huitu and Z. H. Yu, Phys. Lett. B 490 (2000) 87.
\bibitem{abraham}
K. J. Abraham {\em et al.}, preprint hep-ph/0007280.
\bibitem{goity}
J. L. Goity and M. Sher, Phys. Lett. B 346 (1995) 69; erratum
{\em ibid.} B 385 (1996) 500.
\bibitem{masiero}
R. Barbieri and A. Masiero, Nucl. Phys. B 267 (1986) 679.
\bibitem{carlson}
C. E. Carlson, P. Roy and M. Sher, Phys. Lett. B 357 (1995) 99.
\bibitem{white}
B. de Carlos and P. L. White, Phys. Rev. D 55 (1997) 4222.
\bibitem{bhatta}
G. Bhattacharyya, D. Choudhury and K. Sridhar, Phys. Lett. B 355 (1995) 193.
\bibitem{yang}
J. M. Yang, preprint hep-ph/9905486. 
\bibitem{takeuchi}
O. Lebedev, W. Loinaz and T. Takeuchi, Phys. Rev. D 62 (2000) 015003.
\bibitem{keung}
D. Chang and W. Y. Keung, Phys. Lett. B 389 (1996) 294.
\bibitem{pdg}
D. E. Groom {\em et al.} (Particle Data Group), Eur. Phys. Jour. C 15 (2000) 1.
\bibitem{ciuchini}
M. Ciuchini {\em et al.}, JHEP 10 (1998) 008, preprint hep-ph/9808328 v2.
\bibitem{nierste}
S. Herrlich and U. Nierste, Nucl. Phys. B 419 (1994) 292, 
Phys. Rev. D 52 (1995) 6505, Nucl. Phys. B 476 (1996) 27.
\bibitem{dudas}
E. Dudas, S. Pokorski and C. A. Savoy, Phys. Lett. B 356 (1995) 45.
\bibitem{ramond}
P. Bin\'etruy, S. Lavignac and P. Ramond, Nucl. Phys. B 477 (1996) 353.
\bibitem{choi}
E. J. Chun and A. Lukas, Phys. Lett. B387 (1996) 99; 
K. Choi, E. J. Chun and H. Kim, Phys. Lett. B 394 (1997) 89;
K. Choi, E. J. Chun and K. Hwang, Phys. Rev. D 60 (1999) 031301. 
\bibitem{hall}
L. J. Hall and A. Rasin, Phys. Lett. B 315 (1993) 164; 
M. Leurer, Y. Nir and N. Seiberg, Nucl. Phys. B 420 (1994) 468.
\bibitem{haberkane}
H. E. Haber and G. L. Kane, Phys. Rept. 117 (1985) 75.
\bibitem{isajet}
H. Baer {\em et al.}, in {\em Proceedings of the Workshop on Physics at 
Current Accelerators and Supercolliders}, ed. by J. Hewett {\em et al.},
Argonne National Laboratory (1993), preprint hep-ph/9804321. 
\end{thebibliography}
\end{document}